\documentclass[12pt]{article}
\usepackage{amsxtra,amssymb,amsthm,amsmath,latexsym}

\theoremstyle{plain}

\def\R{{\mathbb R}}

\def\oH{{\overset{\circ}{H}}}
\def\oH1{{\overset{\circ}{H}\kern-.02in{}^1}}

\def\bee{\begin{equation*}}
\def\eee{\end{equation*}}
\def\be{\begin{equation}}
\def\ee{\end{equation}}

\begin{document}

\title{On the importance of producing small
impedance particles with prescribed boundary impedance
}

\author{Alexander G. Ramm\\
 Department  of Mathematics, Kansas State University, \\
 Manhattan, KS 66506, USA\\
ramm@math.ksu.edu\\
http://www.math.ksu.edu/\,$\sim$\,ramm
}

\date{}
\maketitle\thispagestyle{empty}

\begin{abstract}
\footnote{PhySC: Condensed matter and materials physics}

The problem of practical preparing small impedance particles with a prescribed boundary impedance is formulated and its importance in physics and technology is discussed. It is shown that if this problem is solved then one can easily prepare materials with a desired refraction coefficient and materials with a desired radiation pattern.

\end{abstract}

\section{Introduction}\label{S:1}
In this section the basic problem of practical 
preparing (producing, manufacturing) of small
impedance particles is discussed. 
Wave scattering by many small impedance particles is developed in \cite{R635}, where
the basic physical assumption is $a\ll d \ll \lambda$. Here $a$ is the characteristic size of the small particles, $d$ is the minimal distance between neighboring particles, and $\lambda$ is the wave length in the medium. We refer for this theory to  \cite{R635}. It is proved there that if one prepares many small particles with prescribed boundary impedance and embed these particles with a specified in  \cite{R635} density in a given material, then one obtains a material whose refraction coefficient approximates any desired refraction coefficient with an arbitrary small error. In particular, one can create meta-materials. How to embed small particles into a given material physicists know. Therefore
the basic practical problem of preparing many small impedance particles with prescribed boundary impedance is of great interest both
technologically and physically.

In section 2 the basic definitions are given. It is explained what a small particle is, what an impedance particle is, what a boundary impedance is.

In section 3 a recipe for creating materials with a desired refraction coefficient is formulated. This recipe is taken from \cite{R635}.   

\section{Basic definitions}\label{S:2}

 Let $D_1\subset \mathbb{R}^3$ be a bounded domain with a connected smooth boundary $S_1$,
 $D_1':=\mathbb{R}^3\setminus D_1$ be the unbounded exterior domain and $S^2$ be the unit sphere in $\mathbb{R}^3$.

Consider the scattering problem:
\be\label{e1}
(\nabla^2+k^2 n_0^2)u=0 \quad in\quad D_1', 
\qquad u_N=\zeta_1 u \quad on \quad S_1, \qquad u=e^{ik\alpha \cdot x}+v,
\ee
where $k>0$ is the wave number, a constant, $\alpha\in S^2$ is a unit vector in the direction of the propagation of the incident plane wave $e^{ik\alpha \cdot x}$,  $N$ is the unit normal to $S_1$ pointing out of $D_1$,
$u_N$ is the normal derivative of $u$, $\zeta_1$
is the boundary impedance, Im$\zeta_1 \le 0$, $n_0>0$ is the refraction coefficient of the small impedance particle $D$, $n_0$ is a constant, 
 the scattered field $v$ satisfies the radiation condition
\be\label{e2} v_r-ikv=o\Big(\frac 1 r\Big), \quad r:=|x|\to \infty.
\ee
The scattering amplitude $A(\beta, \alpha, k)$ is defined by the following formula:
\be\label{e3}
v=A(\beta, \alpha,k)\frac{e^{ikr}}{r} + o\Big(\frac 1 r\Big), \quad r:=|x|\to \infty, \quad \frac {x}{r}=\beta,
\ee
where $\alpha, \beta \in S^2$, $\beta$ is the direction of the scattered wave, $\alpha$ is the direction of the incident wave.
A particle of a charcteristic size $a$ is called small if it is much smaller than the wave length, $kn_0 a\ll 1$.

The function $A(\beta, \alpha, k)$ is called the scattering amplitude.
It is known (see \cite{R190}, p.25, or \cite{R470}) that the solution to the scattering problem \eqref{e1}-\eqref{e3} does exist and is unique. 

If there are many small impedance particles 
$D_j$, embedded in a bounded domain $D$ filled with material
whose refraction coefficient is $n_0$, then the wave scattering problem can be formulated
as follows:
\be\label{e4}
(\nabla^2 +k^2n_0^2)u=0 \quad in \quad D
\ee
\be\label{e5}
u_N=\zeta_m u \quad on \quad S_m, \quad 1\le m \le M, 
\ee
\be\label{e6}
u=u_0(x)+v,
\ee
where the scattered field $v$ satisfies the radiation condition, Im$\zeta_m\le 0$, $M$ is the total number of the embedded particles.
The incident field $u_0$ is assumed known. It satisfies equation \eqref{e4} in $\R^3$, and
as was noted earlier, $n_0=1$ in $D'$. For simplicity we assume that all small particles are of the same characteristic size $a$.

Let $G$ be the Green's function of the scattering problem in the absence of the 
embedded particles. Outside $D$ the refraction coefficient assume to be equal to $1$.Assume that the distribution of small particles is given by the formula
\be\label{e7}
\mathcal{N}(\Delta)=\frac 1 {a^{2-\kappa}}\int_{\Delta} N(x)dx [1+o(1)], \quad a\to 0,
\ee 
where $ \mathcal{N}(\Delta)$ is the number of particles in an arbitrary open subset of $D$,
$N(x)\ge 0$ is a given continuous function,
$\kappa\in [0,1)$ is a number, and the boundary impedance is defined as follows:
\be\label{e8}
\zeta_m=\frac {h_m}{a^{\kappa}}, 
\ee
where Im$h_m\le 0$, $h_m:=h(x_m)$, $x_m\in D_m$
is an arbitrary point and $h(x)$ is a given continuous function in $D$. This function, 
number $\kappa$ and the function $N(x)$ can be chosen by the experimenter.
It is proved in \cite{R635}. p. 48, that the field $u$ in $D$ satisfies, as $a\to 0$, the following integral equation:
 \be\label{e9}
 u(x)=u_0(x)-c_0\int_DG(x,y)h(y)N(y)dy, \quad c_0:=\mathcal{S}/a^2,
 \ee
 where $\mathcal{S}$ is the surface area of a small particle, $h(x)$ and $N(x)$ are defined
in \eqref{e7}-\eqref{e8}. For simplicity we assume here that  $\mathcal{S}$ is the same for all small particles. 
It follows from \eqref{e9} that the new refraction coefficient in $D$, which one gets after embedding many small impedance particles,
is
 \be\label{e10}
n(x)=[ n_0^2-c_0k^{-2}h(x)N(x)]^{1/2}.
\ee
Since $h(x)$ and $N(x)$ are at our disposal,
one can get by formula \eqref{e10} any desirable
refraction coefficient, such that Im$n^2(x)\ge 0$.

Why the equation \eqref{e5} should make sense physically regardless of the size of the particle?

because a problem whose solution exists and is unique must have sense physically.

Why the small impedance particles with a prescribed boundary impedance should exist?

Because the particles with $\zeta =\infty$,
acoustically hard particles, do exist; the 
particles with  $\zeta =0$, acoustically hard
particles, do exist. Small particles with any "intermediate" value of the boundary impedance
should also exist. The problem we raise is
how to produce practically (fabricate) such
particles.

\section{Recipe for creating material with a desired refraction coefficient}\label{S:3}

      Let us formulate the result in the following theorem:
      
        {\bf Theorem 1.} {\em 
            Given $n_0^2(x)$ and a bounded domain $\Omega$, one can create in $\Omega$ a material with
            a desired refraction coefficient\index{refraction coefficient} $n(x), \text{Im} n^2(x) \ge 0$, by embedding
            $M=O(\frac{1}{a^{2-\kappa}})$ small impedance particles according to the distribution
            law \eqref{e7}. The refraction coefficient\index{refraction coefficient} $n_a(x)$, corresponding to a finite $a$,
            approximates the desired refraction coefficient\index{refraction coefficient} $n(x)$ in the sense
   \be\label{e11}          
                \lim_{a\to 0}n_a(x)=n(x).
            \ee
            The functions $h(x)$ and $N(x)$ defined in \eqref{e7}-\eqref{e8} are
            found by the Steps 1, 2 of the Recipe formulated below; see als p. 53 in \cite{R635}.}
        
        Finally, let us discuss briefly the possibility to create material with negative refraction\index{negative refraction} coefficient\index{refraction coefficient}. By formulas (3.1.1) and (3.1.2)
 in \cite{R635}        one gets $n(x) <0$ if the argument of $n_0^2(x)-k^{-2}c_0 N(x)h(x)$ is equal to $2\pi$. Assume that $n_0^2(x) >0$. We know that $k^{-2}c N(x) \ge 0$. Let us take $h(x)=|h(x)|e^{i\varphi}$, where $\varphi>0$ is very small, that is, $\text{Im} h(x) \ge 0$ and $\text{Im} h(x)$ is very small. One proves (see also \cite{R635}, p.52) that equation \eqref{e9} is uniquely solvable for $\text{Im} h \ge 0$ sufficiently small. For such $h(x)$ one has $\arg (n_0^2(x)-k^{-2}c_0 N(x)h(x))$ is very close to $2\pi$ and the square root in \eqref{e10}
  is negative, provided that $\text{Re}(n_0^2(x)-k^{-2}c_0N(x)h(x))>0$. This argument show s that it is possible to create materials with negative refraction coefficient $n(x)$ by embedding in a given material many small particles with properly chosen boundary impedances.

{\bf A recipe for creating materials}
  
  {\bf Problem 1.} Given a material with known $n_0(x)$ in a bounded domain $D$ and $n_0(x)=1$ in $D'$, one wants to create in $D$ a material with a desired $n(x)$.

    {\bf Step 1.}
 Given $n_0^2(x)$ and $n^2(x)$, calculate
   $ p(x):=k^2[n_0^2(x)-n^2(x)]$.
                 This is a trivial step.

 {\bf Step 2.}
   Given $p(x)$ calculate $h(x)$ and $N(x)$ from the equation $p(x)=c_0h(c)N(x)$. The constant $c_0$ one can take to be $c=4\pi$ if the small particles are balls of radius $a$. This we can assume without loss of generality if we are interested in creating materials with a desired refraction coefficient.
   
                        There are infinitely many solutions $h$ and $N$ to the above equation. For example, one can
            fix arbitrarily $N(x)>0$ in $D$, $N(x)=0$ in $D'$, and find $h(x)=h_1(x)+ih_2(x)$
 by the formulas $h_1(x)=\frac{p_1(x)}{4\pi N(x)}$, $ h_2(x)=\frac{p_2(x)}{4\pi N(x)}$.
 Here $p_1(x)=\text{Re} p(x), \quad p_2(x)=\text{Im} p(x)$.
            
            Note that $\text{Im} n^2(x) \ge 0$ implies $\text{Im} h(x) \le 0$, so our assumption $\text{Im} h(x) \le 0$ is satisfied. For example, one may take $N(x)=\text{const} >0$.

 Step 2 is also trivial.

  {\bf Step 3.}
     Given $N(x)$ and $h(x)=h_1(x)+i h_2(x)$, distribute
            $M=O(\frac{1}{a^{2-\kappa}})$ small  impedance balls of radius $a$ in the bounded
             region $D$ according to the distribution law \eqref{e7},
             where $\kappa \in [0,1)$ is the number (parameter) that can be chosen by the
             experimenter.
              Note that condition $d \gg a$ is satisfied automatically
             for the distribution law \eqref{e7}. Indeed, if $d$ is the minimal distance
             between neighboring particles, then there are at most $\frac{1}{d^{3}}$ particles
             in a cube with the unit side, and since $D$ is a bounded domain there are at
             most $O(\frac{1}{d^{3}})$ of small particles in $D$. On the other hand, by
             the distribution law \eqref{e7} one has $\mathcal{N}(D)=O(\frac{1}{a^{2-\kappa}})$.
             Thus $O(\frac{1}{d^{3}})=O(\frac{1}{a^{2-\kappa}})$. Therefore
         $d=O\left(a^\frac{2-\kappa}{3}\right)$.
                      Consequently, condition  $a\ll d \ll \lambda$ is satisfied, $\lambda=\frac {2\pi}{kn_0}$. Moreover
           $ \lim_{a \to 0} \frac{a}{d(a)}=0$.

\end{document}